\documentclass[twocolumn,aps,prl,preprintnumbers]{revtex4}
\usepackage{}
\usepackage{bbm}
\usepackage[latin9]{inputenc}
\usepackage{amsmath}
\usepackage{amssymb}
\usepackage{graphicx}
\usepackage{mathrsfs}
\usepackage{amsfonts}
\usepackage{amsthm}
\usepackage{color}
\usepackage{txfonts}
\usepackage[colorlinks=true,citecolor=blue,linkcolor=blue,urlcolor=blue,anchorcolor=blue]{hyperref}%
\hypersetup{colorlinks=true,citecolor=blue,linkcolor=blue,urlcolor=blue}
\providecommand{\U}[1]{\protect\rule{.1in}{.1in}}
\makeatletter
\@ifundefined{textcolor}{}
{
\definecolor{BLACK}{gray}{0}
\definecolor{WHITE}{gray}{1}
\definecolor{RED}{rgb}{1,0,0}
\definecolor{GREEN}{rgb}{0,1,0}
\definecolor{BLUE}{rgb}{0,0,1}
\definecolor{CYAN}{cmyk}{1,0,0,0}
\definecolor{MAGENTA}{cmyk}{0,1,0,0}
\definecolor{YELLOW}{cmyk}{0,0,1,0}
}
\makeatother
\begin{document}
\title{Twisted magnon as a magnetic tweezer}
\author{Yuanyuan Jiang$^{1}$}
\author{H. Y. Yuan$^{2}$}
\author{Z.-X. Li$^{1}$}
\author{Zhenyu Wang$^{1}$}
\author{H. W. Zhang$^{1}$}
\author{Yunshan Cao$^{1}$}
\author{Peng Yan$^{1}$}
\email[Corresponding author: ]{yan@uestc.edu.cn}
\affiliation{$^{1}$School of Electronic Science and Engineering and State Key Laboratory of Electronic Thin Films and Integrated Devices, University of
Electronic Science and Technology of China, Chengdu 610054, China}
\affiliation{$^{2}$Department of Physics, Southern University of Science
and Technology, Shenzhen 518055, China}

\begin{abstract}
Wave fields with spiral phase dislocations carrying orbital angular momentum (OAM) have been realized in many branches of physics, such as for photons, sound waves, electron beams, and neutrons. However, the OAM states of magnons (spin waves)$-$the building block of modern magnetism$-$and particularly their implications have yet to be addressed. Here, we theoretically investigate the twisted spin-wave generation and propagation in magnetic nanocylinders. The OAM nature of magnons is uncovered by showing that the spin-wave eigenmode is also the eigenstate of the OAM operator in the confined geometry. Inspired by optical tweezers, we predict an exotic ``magnetic tweezer" effect by showing skyrmion gyrations under twisted magnons in exchange-coupled nanocylinder$|$nanodisk heterostructure, as a practical demonstration of magnonic OAM transfer to manipulate topological spin defects. Our study paves the way for the emerging magnetic manipulations by harnessing the OAM degree of freedom of magnons.
\end{abstract}

\maketitle
In the past decades, the quantized orbital angular momentum (OAM) of wave fields with spatially twisted phase structure has been widely investigated, ranging from photons \cite{Allen1992,Allen2003,Torres2011,Andrews2012,Mair2001,Vallone2014,Furhapter2005,Tamburini2011,Grier2003,Jones2015,Yang2018}, electron beams \cite{Bliokh2007,Uchida2010,Verbeeck2010,McMorran2011,Mafakheri2017,Silenko2017,Lloyd2017}, and acoustic waves \cite{Nye1974,Hefner1999,Anhauser2012,Hong2015,Marzo2018,Baresch2018,Zhang2018} to neutrons \cite{Clark2015,Cappelletti2018} and gluons \cite{Ji2017}. The OAM is associated with rotation of a (quasi-)particle about a fixed axis, and is encoded in the spiral phase profile of the particle's wavefunction characterized by an azimuthal $\phi$ phase dependence $e^{i\ell\phi}$ with a nonzero topological charge $\ell$ (an integer) and a vanishing field at the core. The OAM component in the direction of rotational axis has the quantized value $\ell\hbar$ with $\hbar$ the reduced Planck constant, in contrast to the spin angular momentum (SAM) originating from the wave polarization. Such twisted OAM states have a phase dislocation on the axis that is sometimes referred to as a vortex. Vortices with high OAM can be achieved using spiral phase plates, computer-generated holograms, mode conversions, and spatial modulators, among others \cite{Verbeeck2010,McMorran2011,Lloyd2017,Cai2012,Fickler2016}. However, the OAM state of magnons (or spin waves)$-$as elementary excitations in ordered magnets$-$has received little attention by the community \cite{Tsukernik1966,Yan2013} and its practical implication has never been addressed, although their linear momentum and SAM have been extensively explored in the context of Brillouin light scattering spectroscopy \cite{Demokritov2001}, magnon-driven dynamics of topological spin texture \cite{Yan2011,Jiang2013,Wang2012,Wang2015,Lin2013}, Bose-Einstein magnon condensation \cite{Demokritov2006}, etc.

\begin{figure}
  \centering
  \includegraphics[width=0.48\textwidth]{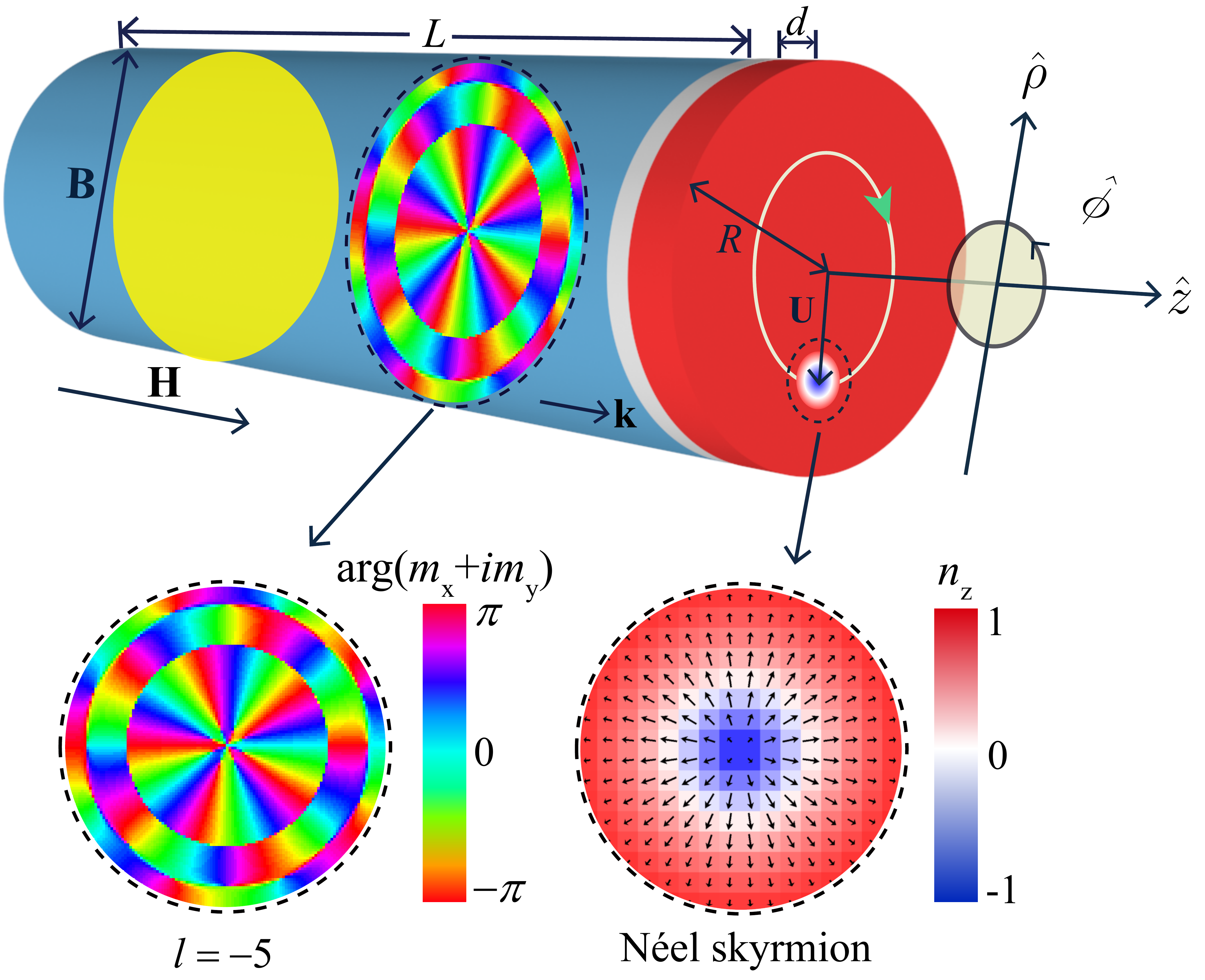}\\
  \caption{Schematic illustration of a heterostructured nanocylinder exchange-coupled to a chiral magnetic nanodisk hosting a N\'{e}el-type skyrmion. A static field $\textbf{H}$ is applied along the $z$-direction. A spin-wave beam with the wavevector $\mathbf{k}$ and OAM quantum number $\ell=-5$ is excited by a local microwave field $\textbf{B}$ applied over the yellow-disk region, leading to a steady skyrmion gyration around the disk center.}\label{fig1}
\end{figure}

In this Letter, we uncover the OAM nature of magnons by studying the spin-wave dynamics in magnetic nanocylinders (see Fig. \ref{fig1}). As a proof of concept, we generate magnon beams carrying OAM quantum number $|\ell|\leqslant8$ via localized spatiotemporal fields. Inspired by the notion of optical tweezers \cite{Ashkin1986}, we predict a ``magnetic tweezer" effect by demonstrating twisted-magnon induced skyrmion gyration in a chiral magnetic nanodisk exchange-coupled to one end of the magnetic nanocylinder, as a practical application of the magnonic OAM transfer to matter.

We start with the following Hamiltonian modelling a uniaxial ferromagnet of the cylindrical geometry,
\begin{equation}\label{Hamiltonian}
  \mathcal{H}=\int d\textbf{r} \bigg[\frac{A}{M^{2}_{s}} {(\nabla \textbf{M})}^{2}-\mu_{0}\textbf{M}\cdot(\textbf{H}+\textbf{h})\bigg],
\end{equation}
where $\textbf{M}=M_{s}\textbf{m}$ is the local magnetization with the saturated value $M_{s}$ and the direction $\textbf{m}$, $A$ is the exchange constant, $\textbf{H}=H_{0}\hat{z}$ is the external field along the $z$-axis (the symmetry axis of the cylinder), $\mu_{0}$ is the vacuum permeability, and $\textbf{h}$ is the dipolar field satisfying the magnetostatic equation $\nabla\times\textbf{h}(\textbf{r},t)=0$ and $\nabla\cdot\big[\textbf{h}(\textbf{r},t)+M_{s}\textbf{m}(\textbf{r},t)\big]=0$. We therefore have $\textbf{h}(\textbf{r},t)=-\nabla\Phi(\textbf{r},t)$ where $\Phi$ is a magnetostatic potential. The spatiotemporal evolution of magnetization is governed by the Landau-Lifshitz-Gilbert (LLG) equation,
\begin{equation}\label{LLG}
  \frac{\partial\textbf{m}}{\partial t}=-\gamma \mu_{0}\textbf{m}\times \textbf{H}_{\text{eff}}+\alpha\textbf{m}\times\frac{\partial\textbf{m}}{\partial t},
\end{equation}
where $\gamma$ is the gyromagnetic ratio, $\alpha$ is the Gilbert damping, and $\textbf{H}_{\text{eff}}=-\mu^{-1}_{0}\delta \mathcal{H}/\delta \textbf{M}=H_{0}\hat{z}+\textbf{h}(\textbf{r},t)+\frac{2A}{\mu_{0}M_{s}}\nabla^{2}\textbf{m}(\textbf{r},t)$ is the effective field. We consider the spin-wave excitation on top of a uniform magnetization $\textbf{m}=(m_{x},m_{y},1)$ with $m^{2}_{x}+m^{2}_{y}\ll 1$, and pursue the time-harmonic solution $m_{x(y)}(\textbf{r},t)=m_{x(y)}(\textbf{r})e^{-i\omega t}$ and $\Phi(\textbf{r},t)=\Phi(\textbf{r})e^{-i\omega t}$ with the frequency $\omega$. Substituting these terms into the coupled magnetostatic and LLG equations and adopting the linear approximation, we obtain
\begin{subequations}\label{Linearize}
\begin{eqnarray}
  i\bar{\omega} m_{x}&=&(H_{0}-\bar{A}\nabla^{2})m_{y}+\frac{\partial\Phi}{\partial y},\\
  -i\bar{\omega} m_{y}&=&(H_{0}-\bar{A}\nabla^{2})m_{x}+\frac{\partial\Phi}{\partial x},\\
  \nabla^{2}\Phi &=&M_{s}(\frac{\partial m_{x}}{\partial x}+\frac{\partial m_{y}}{\partial y}),
  \end{eqnarray}
\end{subequations}
with $\bar{\omega}=\omega/(\gamma \mu_{0})$ and $\bar{A}=2A/(\mu_{0}M_{s})$. The magnetic potential within the cylinder takes the form $\Phi(\rho,\phi,z)\sim J_{n}(\kappa\rho)e^{in\phi+ikz}$. Here, $J_{n}(\kappa\rho)$ is the Bessel function of the first kind, $n=0,\pm 1, \pm 2,\ldots$ is the azimuthal quantum number, $k$ is the longitudinal wave number, and $\kappa$ is the transverse wave number. Substituting the Bessel profile into Eqs. \eqref{Linearize} leads to the following dispersion relation: $-H_{0}M_{s}k^{2}+\big[H_{0}(H_{0}+M_{s})-\bar{\omega}^{2}-\bar{A}M_{s}k^{2}\big](\kappa^{2}+k^{2})+\bar{A}(2H_{0}+M_{s})(\kappa^{2}+k^{2})^{2}+\bar{A}^{2}(\kappa^{2}+k^{2})^{3}=0$
which is cubic in $\kappa^{2}$, so that for each combination of $n$ and $k$, we have three linearly independent solutions of the magnetic potential $\Phi(\rho,\phi,z)=\sum^{3}_{j=1}c_{j}J_{n}(\kappa_{j}\rho)e^{i(n\phi+kz)}$ for $\rho\leq R$ and $\Phi(\rho,\phi,z)=c_{4}K_{n}(k\rho)e^{i(n\phi+kz)}$ for $\rho>R$, where $R$ is the radius of the nanocylinder and $K_{n}(k\rho)$ is the modified Bessel function of the second kind. Accordingly, the radial and azimuthal components of the dynamical magnetization are given by
$m_{\rho}(\rho,\phi,z)=\frac{1}{2}\sum^{3}_{j}c_{j}\kappa_{j}\bigg[\frac{J_{n+1}(\kappa_{j}\rho)}{\bar{A}(\kappa^{2}_{j}+k^{2})+H_{0}+\bar{\omega}}-\frac{J_{n-1}(\kappa_{j}\rho)}{\bar{A}(\kappa^{2}_{j}+k^{2})+H_{0}-\bar{\omega}}\bigg]e^{i(n\phi+kz)},
$ $m_{\phi}(\rho,\phi,z)=\frac{-i}{2}\sum^{3}_{j}c_{j}\kappa_{j}\bigg[\frac{J_{n+1}(\kappa_{j}\rho)}{\bar{A}(\kappa^{2}_{j}+k^{2})+H_{0}+\bar{\omega}}+\frac{J_{n-1}(\kappa_{j}\rho)}{\bar{A}(\kappa^{2}_{j}+k^{2})+H_{0}-\bar{\omega}}\bigg]e^{i(n\phi+kz)},
$
in which we have defined $m_{\rho}=\cos\phi m_{x}+\sin\phi m_{y}$ and $m_{\phi}=-\sin\phi m_{x}+\cos\phi m_{y}$. From Noether's theorem, one can derive the OAM of twisted magnons
\begin{equation}\label{OAMz}
  \mathcal {J}_{z}=\frac{\hbar}{S}\int \!\!\! \int{(\nabla\psi\times\mathbf{r})_{z}dxdy},
\end{equation}where $S=\pi R^{2}$ is the cross-section area of the nanocylinder and $\psi(\mathbf{r})=\arg(m_{x}+im_{y})$ \cite{Yan2013}. Substituting the dynamical magnetization profile into \eqref{OAMz}, we obtain $\mathcal {J}_{z}=\ell\hbar$ with $\ell=n-1$, the quantized OAM per twisted magnon.

The determination of the four unknown parameters $c_{1},c_{2},c_{3},$ and $c_{4}$ depends on the boundary conditions on the nanocylinder surface, i.e., the continuity of $\Phi$, the normal components
of the magnetic induction and the magnetization \cite{notesp}. This allows us writing the condition for the existence of nontrivial solutions
\begin{equation}\label{Determinant}
  \det \Delta(\bar{\omega},\kappa_{1},\kappa_{2},\kappa_{3})=0,
\end{equation}
where $\Delta(\bar{\omega},\kappa_{j})$ is the $4\times4$ coefficient matrix of boundary conditions \cite{notesp}. We shall solve the problem numerically by searching for zeros of $\Delta$ \cite{Sattler2011}.
\begin{figure}
  \centering
  \includegraphics[width=0.48\textwidth]{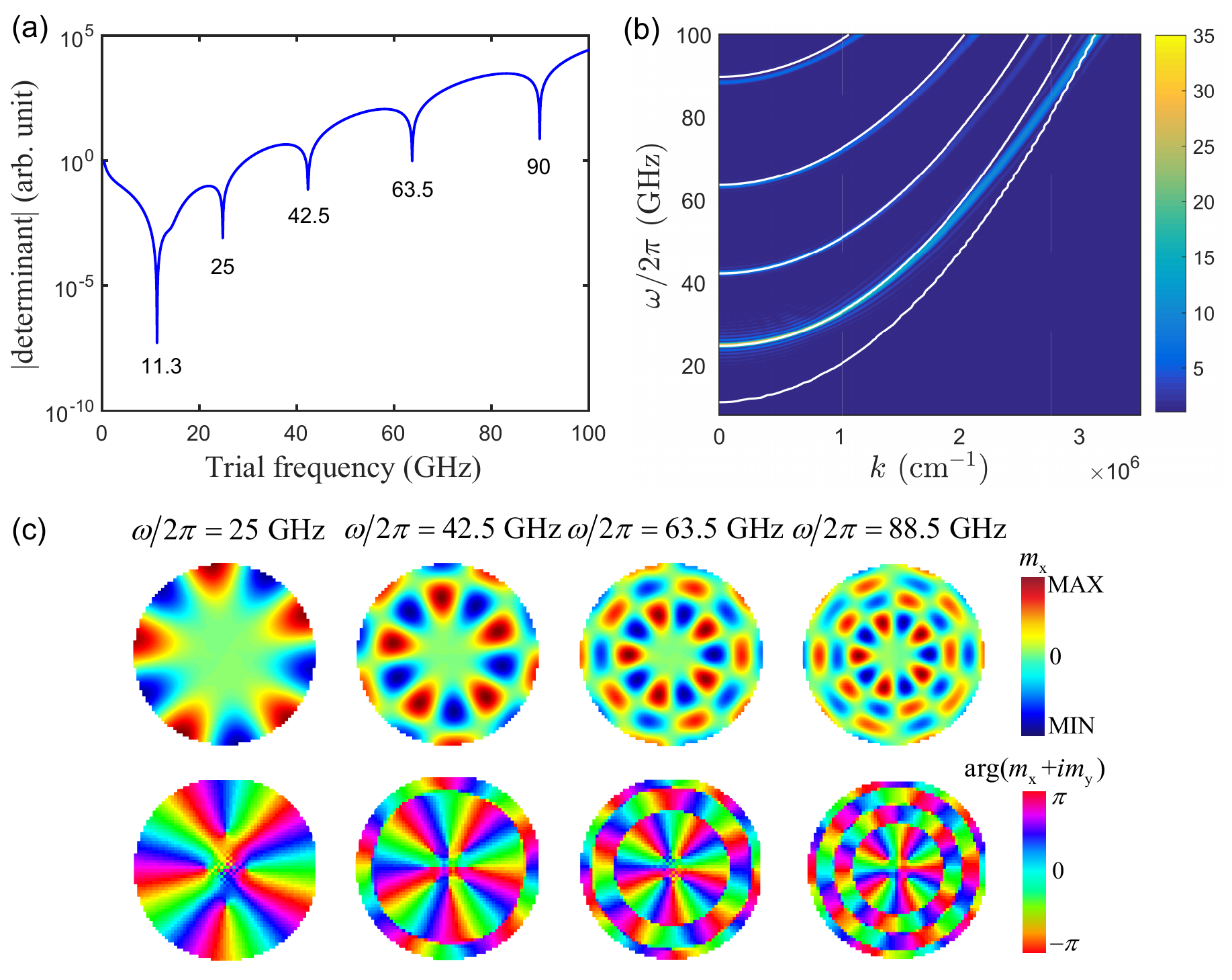}\\
  \caption{(a) Boundary-value determinant versus trial frequencies for a longitudinal wavevector $k=1\times10^{5}$ cm$^{-1}$ and $\ell=-5$. (b) Spin-wave spectrum in the nanocylinder. Solid curves are theoretical results, comparing with the FFT transform amplitude of $m_x$ from micromagnetic simulations. (c) Cross-sectional distribution of spin-wave beams with $0,1,2,$ and $3$ radial nodes (from left to right) \cite{Note1}.}\label{fig2}
\end{figure}

We consider an isolated yttrium iron garnet (YIG) nanocylinder of length $L=2$ $\mu$m and radius $R=60$ nm \cite{notesp}. The external field is $\mu_{0}H_{0}=0.4$ T. We consider the case $\ell=-5$. In Fig. \ref{fig2}(a), we show the magnitude of the determinant $\Delta$ as a function of trial frequencies for $k=1\times10^{5}$ cm$^{-1}$. Below $100$ GHz, we find five frequencies at $11.3,$ $25,$ $42.5,$ $63.5,$ and $90$ GHz satisfying boundary conditions. Full band structure is plotted by white curves in Fig. \ref{fig2}(b). To verify the theoretical spin-wave spectrum, we simulate the LLG equation \eqref{LLG} using MuMax3 \cite{Vansteenkiste2014}. To excite higher-order spin-wave modes, we apply a microwave field and analyze the Fourier transformation (FFT) of the spatiotemporal oscillation of $m_{x}$, as shown in Fig. \ref{fig2}(b) \cite{notesp}. Simulation results compare well with theoretical curves, except the lowest band which represents the cross-sectionally uniform spin-wave mode and cannot be generated by an inhomogeneous microwave field. Figure \ref{fig2}(c) shows the cross-sectional distribution of different modes emerging in Fig. \ref{fig2}(a). We find that these modes actually have $0,1,2,$ and $3$ radial nodes, viewed from left to right, respectively. The magnonic potential well between two neighboring nodes can be utilized to trap the topological spin defects. However, only when its width is comparable to or larger than the size of the defects, the trapping is efficient. We thus focus on the twisted magnons with $2$ nodes [the 2nd white curve from the top in Fig. \ref{fig2}(b)].

To illustrate how twisted magnons can manipulate magnetic textures, we consider a heterostructure shown in Fig. \ref{fig1}. The nanodisk is a chiral ferromagnet containing a stabilized skyrmion. Twisted magnons are excited in the YIG nanocylinder and then propagate rightward $(+\hat{z})$ to interact with the skyrmion in the nanodisk. Here, the magnetization dynamics in the nanodisk is described by LLG equation as
\begin{equation}
\frac{\partial\textbf{n}}{\partial t}= -\gamma\mu_{0}\mathbf{n}\times (\mathbf{H}'_{\text{eff}}+\mathbf{H}_{\text{ex}})+\alpha'\mathbf{n}\times\frac{\partial\textbf{n}}{\partial t},\\
\label{llg}
\end{equation}
where $\mathbf{n}$ is the normalized magnetization in the nanodisk, $\mathbf{H}'_{\text{eff}}$ is the effective field, $\mathbf{H}_{\text{ex}}$ is an interfacial field resulting from the nanocylinder$|$nanodisk exchange coupling, and $\alpha'$ is the Gilbert damping. Generally, the exchange coupling $E_{\text{ex}}=-J_{\text{ex}} \mathbf{n}(\textbf{r},t) \cdot \mathbf{m}(\textbf{r},t)$ such that $\mathbf{H}_{\text{ex}}=J_{\text{ex}} \mathbf{m}(\textbf{r},t)/(\mu_{0}M'_{s})$ with $M'_{s}$ being the saturation magnetization of the nanodisk. Via the interfacial exchange coupling, twisted magnons can transfer their OAM to skyrmions. Below we address this idea analytically and numerically.

We adopt Thiele's model to make analytical predictions. By performing $[\mathbf{n} \times \text{Eq.}\ (\ref{llg})]\cdot \nabla \mathbf{n}$
and considering the collective coordinate $\mathbf{n}(\mathbf{r},t)=\mathbf{n}[\mathbf{r}-\mathbf{U}(t)]$, we obtain
\begin{equation}\label{Thiele}
\mathcal{G}\hat{z} \times \frac{d\mathbf{U}}{dt}-\alpha'\mathcal{D} \cdot \frac{d\mathbf{U}}{dt}+\mathbf{F}=0,
\end{equation}
where $\mathbf{U}$ is the displacement of the skyrmion core from the disk center; $\mathcal{G}=-4\pi$$Qd M'_{s}$/$\gamma$ is the gyroscopic constant with $Q=\frac{1}{4\pi}\int \!\!\! \int{\mathbf{n}\cdot(\frac {\partial \mathbf{n}}{\partial {x} } \times \frac {\partial \mathbf{n}}{\partial y } )dxdy}$ being the topological charge [$Q=-1$ for the skyrmion in Fig. \ref{fig1}]; $d$ is the thickness of the nanodisk; and $\mathcal{D}_{ij}=(d M'_{s}$/$\gamma)\int \!\!\! \int{\partial_i \mathbf{n} \cdot \partial_j \mathbf{n}dxdy}$ is the dissipation tensor. The driving force includes two parts $\mathbf{F}=\mathbf{f}+\mathbf{g}$, where $\mathbf{f}(t)=-\mathcal{K}_{0}\textbf{U}(t)$ is the restoring force from disk boundary with the positive spring constant $\mathcal{K}_{0}$ \cite{Yang2018B}, and $g_i(\textbf{U}(t),t)=J_{\text{ex}}d \int \!\!\! \int{ \mathbf{m}(\textbf{r},t)\cdot \partial_i\mathbf{n}[\textbf{r}-\textbf{U}(t)] dxdy}$ is the force due to twisted magnons from the nanocylinder. Typically, the skyrmion dynamics is much slower than the spin-wave precession, i.e., $|\dot{\textbf{U}}(t)|/U(t)\ll|\dot{\mathbf{m}}|$ with $U(t)=|\mathbf{U}(t)|$. The $\mathbf{g}$ force can thus be divided into two terms $\mathbf{g}=\mathbf{g}'[\mathbf{U}(t)]+\mathbf{g}''(t)$. Here, $\mathbf{g}'$ is the slow component with the same time-scale as the skyrmion guiding center, whereas $\mathbf{g}''$ evolves as fast as the twisted magnons and $\langle\mathbf{g}''\rangle=0$ by coarse-graining. Considering the leading-order terms only, we can expand the total driving-force as
\begin{equation}\label{Drivingforce}
  \mathbf{F}=-\mathcal{K}\mathbf{U}(t)-\lambda\hat{\phi},
\end{equation}
where $\mathcal{K}=\mathcal{K}_{0}+\Delta\mathcal{K}$ is the effective spring constant with $\Delta\mathcal{K}$ contributed by twisted magnons and $\lambda$ is the coefficient of azimuthal force. In a large-sized nanodisk, $\Delta\mathcal{K}$ can dominate over $\mathcal{K}_{0}$. By seeking the steady-state solution for skyrmion gyrations, i.e., $U=\text{const.}$, and substituting \eqref{Drivingforce} into \eqref{Thiele}, we obtain
\begin{equation}\label{Solution}
  \dot{\mathbf{U}}(t)=-\frac{\mathcal{K}U}{\mathcal{G}}\hat{\phi},\ \ \text{and  }U=\frac{\lambda\mathcal{G}}{\alpha'\mathcal{K}\mathcal{D}}.
\end{equation}
\begin{figure}[ptbh]
\begin{centering}
\includegraphics[width=0.48\textwidth]{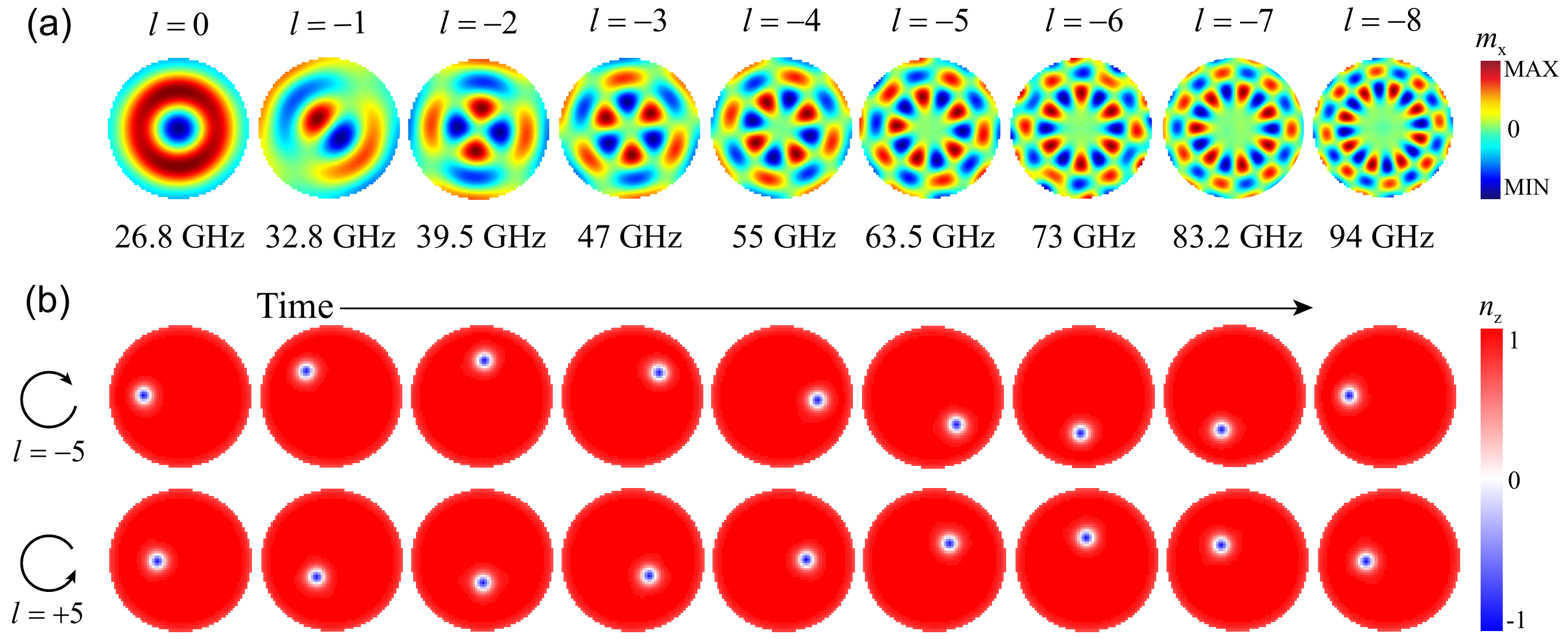}
\par\end{centering}
\caption{(a) Cross-sectional view of spin-wave beams carrying different OAMs. (b) Time-evolution of N\'{e}el skyrmions driven by twisted magnons with OAM $\ell=\mp5$. The time intervals between successive snapshots are $1.65$ ns and $1.98$ ns, respectively. A negative (positive) OAM induces a clockwise (counter-clockwise) skyrmion gyration.}\label{fig3}
\end{figure}
The above result uncovers the two-fold contributions of twisted magnon: First, it induces an extra confining force which determines the gyration frequency of the circular motion. Second, it generates an azimuthal force to compensate the dissipation to sustain a stable gyration and prevent the skyrmion from falling into the disk center. Under a suitable orbit radius $U$, these two physical processes reach a balance. Because the spring model with constant coefficients is exact only when $U/R\ll 1$ \cite{Yang2018B}, both $\mathcal{K}$ and $\lambda$ may depend on the orbit radius $U$ [see Fig. \ref{fig4}(c)]. Below we perform full micromagnetic simulations \cite{Vansteenkiste2014} to verify our theoretical predictions.

We consider a Co$/$Pt nanodisk exchanged-coupled to the YIG cylinder \cite{notesp}. The Laguerre-Gaussian (LG) excitation microwave field $\textbf{B}_{\ell}(\rho,\phi,t)=B_{0}(\frac{\rho}{w})^{|\ell|}e^{-\frac{\rho^{2}}{w^{2}}}\mathcal {L}^{|\ell|}_{2}(\frac{2\rho^{2}}{w^{2}})\cos(-\omega t+\ell\phi)\hat{x}$ is locally applied at $z=-20$ nm (we set the interface as $z=0$). Here, $w$ is the width of the beam waist, $B_{0}$ is the field amplitude, and $\mathcal {L}^{|\ell|}_{2}$ is the generalized Laguerre function. We have numerically generated twisted $2$-node magnon beams with a wavevector $k=1\times10^{5}$ cm$^{-1}$ and  $-8\leqslant\ell\leqslant0$  [see Fig. \ref{fig3}(a)] \cite{notesp}.

Figure \ref{fig3}(b) shows the steady skyrmion gyration under twisted magnon beams. The initial position of skyrmion is $\mathbf{U}(t=0)=(-22, 0)$ nm [see Fig. \ref{fig4}(a)]. For $\ell=-5$, we observe a clockwise skyrmion circular motion with an average velocity $15.2$ m$/$s along the orbit of radius $31.9$ nm [shown in Figs. \ref{fig3}(b) and \ref{fig4}(a)]. To prove that the rotational motion is indeed due to the OAM transfer from twisted magnons to the skyrmion, we reverse the OAM of the magnon beam to $\ell=+5$ without changing the rest parameters. We then observe a counter-clockwise gyration of the skyrmion, as expected. However, the skyrmion velocity is now significantly reduced to $7.82$ m$/$s along a shrunk orbit of a radius $19.7$ nm [see Figs. \ref{fig3}(b) and \ref{fig4}(c)]. The symmetry breaking of $\ell=-5$ and $\ell=+5$ is due to the topological nature of the skyrmion. Given a skyrmion profile with a particular winding number ($Q=-1$ in our case), it has an intrinsic clockwise gyration [Fig. \ref{fig4}(a)]. On the other hand, twisted magnons carrying $\ell=-5$ and $\ell=+5$ tend to induce skyrmion rotation in clockwise and counter-clockwise direction, respectively. Combing these effects, the intrinsic topology of skyrmion tends to boost (hinder) its rotation by twisted magnons with $\ell=-5$ ($\ell=+5$). As a comparison, we consider the case with a vanishing OAM $(\ell=0)$ and find the skyrmion falling into the disk center [gray curve in Fig. \ref{fig4}(a)], consistent with the OAM transfer picture. We also perform simulations without magnons and find that the skyrmion naturally gyrates into the disk center but with a much longer time.

\begin{figure}
  \centering
  \includegraphics[width=0.48\textwidth]{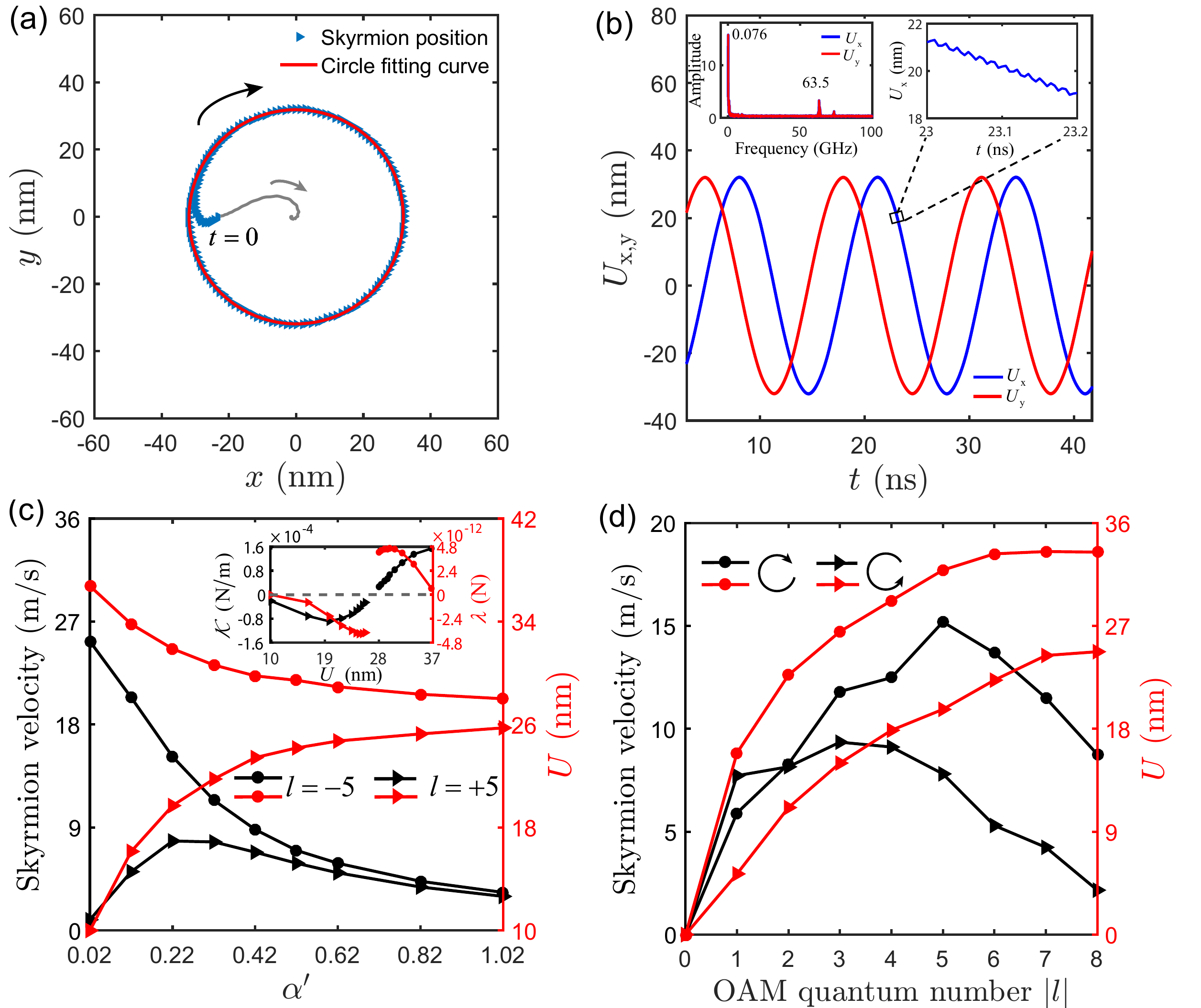}\\
  \caption{(a) Skyrmion gyration path (symbols) and the circle fitting (red circle) for $\ell=-5$. Gray curve represents the path for $\ell=0$. (b) Time-dependence of $U_{x}$ (blue line) and $U_{y}$ (red line). Inset shows the high-frequency details of the FFT spectrum (left) and $U_{x}$ (right). Skyrmion velocity and orbit radius as a function of (c) damping parameter $\alpha'$ and (d) OAM quantum number $\ell$. Inset plots the $U$-dependence of $\mathcal{K}$ and $\lambda$.}\label{fig4}
\end{figure}

Figure \ref{fig4}(b) plots the time-dependence of the guiding center $U_{x,y}$ driven by the spin-wave beam with $\ell=-5$, indicating a period of $13.2$ ns. However, by carefully investigating the orbit, we observe a fast oscillation mode (right inset). Through the FFT analysis of $\mathbf{U}(t)$, we identify two peaks at 0.076 and 63.5 GHz (left inset), which correspond to the skyrmion gyration and spin-wave precession, respectively. The relatively lower peak of the fast mode justifies our coarse-graining treatment on $\mathbf{g}''$. Furthermore, we confirm the important role of damping. As shown in Fig. \ref{fig4}(c), the skyrmion velocity monotonically decreases with $\alpha'$ for $\ell=-5$. However, for $\ell=+5$ the skyrmion velocity first increases until a certain value and then decreases. We also find a shrinking (surprisingly expanding) of the gyration orbit for $\ell=-5$ ($\ell=+5$) when $\alpha'$ increases. To quantitatively interpret these numerical observations, we derive the $U$-dependence of parameters $\mathcal{K}$ and $\lambda$ in Eq. \eqref{Drivingforce} (see the inset), which shows a sign reversal when $\ell$ is switched from $-5$ to $+5$. The orbit shrinking and expanding can thus be understood as follows: For $\ell=-5$, the spring coefficient $\mathcal{K}>0$ and the radial force in Eq. \eqref{Drivingforce} is centripetal. An increased damping will slow down skyrmion's gyrating velocity and shrink its orbit radius. For $\ell=+5$, the spring coefficient $\mathcal{K}<0$ and the radial force becomes centrifugal. However, the Thiele's equation \eqref{Thiele} is invariant by transformations $\mathcal{G}\rightarrow-\mathcal{G},\lambda\rightarrow-\lambda,\mathcal{K}\rightarrow-\mathcal{K},$ and $\alpha'\rightarrow-\alpha'$, through which the radial force recovers its centripetal nature. The $-\alpha'$ can be viewed as a gain \cite{Yang2018L}, instead of damping. An increased $\alpha'$ thus represents an enhanced gain, which naturally leads to the expansion of gyration orbits. The non-monotonic $|\ell|$-dependence of the skyrmion velocity indicates the possibility to optimize twisted spin-wave beams and material parameters, although a larger OAM magnitude always expands the gyration orbit [Fig. \ref{fig4}(d)].

In numerical simulations, we use a LG microwave field confined inside the magnetic cylinder purely as a computational method to generate twisted magnons, rather than a practical proposal to induce gyration of skyrmion, although it could induce a faster skyrmion gyration when it is directly applied on the Co$/$Pt disk. From an experimental point of view, spin-wave beams carrying various OAM can be excited by Brillouin light scattering \cite{Haigh2018} or by magnetic spiral phase plates \cite{Jia2019}. Very recently, twisted photons are proposed to drive a rotational motion of skyrmion, which however suffers from the diffraction limit \cite{Yang2018,Fujita2016,Fujita2017}. In contrast to its photonic counterpart, the magnonic vortex matches all time- and length-scales of skyrmion dynamics, free from the mentioned limitations. Furthermore, our magnonic OAM induced skyrmion velocity can reach tens of meters per second with a low power consumption. It is comparable to the current-driven skyrmion velocity in nanodisks that, however, work at very high electric-current densities \cite{SanchezNJP2016,ZhangNJP2015}. The magnonic tweezers can also find applications in manipulating other types of spin defects, such as magnetic vortex \cite{Li2019}, bobber \cite{Rybakov2015,Zheng2018}, meron \cite{Yu2018}, hopfion \cite{Dai2018,Liu2018,Sutcliffe2018,Rybakov2018,Wang2019}, etc. Twisted magnons in antiferromagnets \cite{Tsukernik1968} may manifest themselves THz magnetic tweezers. Pushing the magnonic OAM manipulation into quantum regions \cite{Huaiyang2018,Huaiyang2020} is also an interesting issue.

In summary, we predicted magnetic tweezers effect by demonstrating that magnonic vortices can drive the dynamics of topological spin defects. Taking the magnetic skyrmion as an example, we showed both theoretically and numerically that twisted spin-wave beams can induce a steady-state skyrmion gyration in a hybrid nanocylinder$|$nanodisk structure. Our work opens the door for twisted magnonics, an emerging field about all-magnonic manipulation by harnessing the OAM of magnons, in addition to their linear momentum and SAM degrees of freedom.

\begin{acknowledgments}
This work was funded by the National Natural Science Foundation of China (Grants No. 11604041, No. 11704060, and No. 11904048) and the National Key Research Development Program under Contract No. 2016YFA0300801. H.Y. acknowledges financial support from the National Natural Science Foundation of China under Grant No. 61704071.

Y.J. and H.Y. contributed equally to this work.
\end{acknowledgments}

\end{document}